\documentclass[12pt]{article}
\usepackage{amsmath}
\usepackage{graphicx}
\usepackage{enumerate}
\usepackage{url} 

\newcommand{\blind}{1}

\usepackage[normalem]{ulem}

\usepackage{amssymb}
\usepackage{amsmath,amsthm, mathtools,commath}
\usepackage{graphics, color}
\usepackage{booktabs}
\usepackage{arydshln}

\usepackage{graphicx}
\usepackage{epsfig}
\usepackage{makeidx}
\usepackage{kotex}
\usepackage{fullpage}
\usepackage{comment}
\usepackage{bigints}

\usepackage[round]{natbib}
\newcommand{\T}{\top}

\newcommand{\biblist}{\begin{list}{}
{\listparindent 0.0cm \leftmargin 0.50cm \itemindent -0.50 cm
\labelwidth 0 cm \labelsep 0.50 cm
\usecounter{list}}\clubpenalty4000\widowpenalty4000}
\newcommand{\ebiblist}{\end{list}}

\newcommand{\bx}{\bm{x}}

\newcommand{\bz}{\bm{z}}

\newcommand{\bbeta}{{\mbox{\boldmath$\beta$}}}
\newcommand{\blambda}{{\mbox{\boldmath$\lambda$}}}

\newcommand{\bomega}{{\mbox{\boldmath$\omega$}}}
\newcommand{\bgamma}{{\mbox{\boldmath$\gamma$}}}

\newcommand{\bphi}{{\mbox{\boldmath$\phi$}}}

\newtheorem{theorem}{Theorem}

\newtheorem{remark}{Remark}

\newcommand{\Norm}[1]{\left\Vert#1\right\Vert}

\usepackage{xcolor}

\usepackage{graphicx, graphics, color, subcaption}

\usepackage{amsmath, amssymb, amsfonts, amsthm}
\usepackage{mathtools, mathrsfs, bbm, bm, commath}

\DeclareMathOperator*{\argmax}{arg\,max}
\DeclareMathOperator*{\argmin}{arg\,min}

\usepackage{float, multirow}

\usepackage{comment}

\usepackage{enumitem} 

\usepackage{xr}

\numberwithin{equation}{section}

\renewcommand{\theequation}{\thesection.\arabic{equation}}

\begin{document}
\def\spacingset#1{\renewcommand{\baselinestretch}%
{#1}\small\normalsize} 
\spacingset{1}

\if1\blind
{
  \title{\bf 
  
  Generalized entropy calibration for analyzing voluntary survey data  }
  \author{Yonghyun Kwon \\
  Department of Mathematics, Korea Military Academy, Seoul, Republic of Korea \\
  yhkwon@kma.ac.kr\\
  and \\
   Jae Kwang Kim \\
  Department of Statistics, Iowa State University, Iowa, USA\\
  jkim@iastate.edu \\  
  and \\
  Yumou Qiu \\
  School of Mathematical Sciences, Peking University, Beijing, China \\ qiuyumou@math.pku.edu.cn  }
\date{}
\maketitle
} \fi

\if0\blind
{
 \bigskip
 \bigskip
 \bigskip
 \begin{center}
   {\LARGE\bf Generalized entropy calibration for analyzing voluntary survey data 
     }
\end{center}
 \medskip
} \fi

\bigskip




\begin{abstract}
Statistical analysis of voluntary survey data is an important area of research in survey sampling. We consider a unified approach to voluntary survey data analysis under the assumption that the sampling mechanism is ignorable. Generalized entropy calibration  is introduced as a unified tool for calibration weighting to control the selection bias. We first establish the relationship between the generalized calibration weighting and its
dual expression for regression estimation. The dual relationship is critical in identifying the
implied regression model and developing model selection for calibration weighting. Also,
if a linear regression model for an important study variable is available, then two-step calibration method can be used to smooth the final weights and achieve the statistical efficiency. Asymptotic properties of the proposed estimator  are investigated. 
Results from a limited simulation study are also presented. The proposed method is applied to a real data application in the context of data integration. 
\end{abstract}

\noindent%
\vfill
\newpage 

\spacingset{1.9} 

\section{Introduction}
\label{intro}

While probability samples serve as the gold standard in social science and related fields to estimate population quantities and monitor policy effects, their acquisition and analysis becomes challenging due to their high cost and low response rates \citep{williams2018trends,kalton2019developments}. During the past 20 years, non-probability samples, on the other hand, have been increasingly prevailing. However, non-probability samples suffer from selection bias because of unknown selection probabilities, and it may lead to erroneous inference if such selection bias is overlooked \citep{couper2000web,bethlehem2016solving,elliott2017inference,meng2018statistical}. Consequently, addressing selection bias in non-probability samples constitutes a significant research area.

Voluntary samples are typically collected through a process in which individuals choose to participate on their own. Because the sample inclusion probabilities are unknown, statistical inference for voluntary samples is challenging. To adjust for selection bias in non-probability samples, many studies assume that a set of auxiliary variables, or their population totals, is available either at the population level or through a probability-based reference sample with known sampling weights. Following this approach, we consider the scenario where auxiliary variables, or their population totals, are available at the population level.

Existing approaches to handling nonprobability samples can be classified into two types. The first is the propensity score approach, which relies on modeling the selection propensity score function. In this approach, the selection probability for each unit in the nonprobability sample is estimated, and the inverse of this probability is applied to each unit. \citet{elliott2017inference} proposed estimating the inclusion probability by combining probability and nonprobability samples. For additional references, see \citet{valliant2020comparing, liu2023correcting}. \citet{chen2020doubly, chen2023dealing} explored the calibration weighting approach, which leads to doubly robust estimation.


The second approach is the prediction approach based on regression modeling for the study variable of interest. Using the outcome regression model, the unobserved variable can be imputed—a process also known as mass imputation. A theoretical framework for mass imputation in nonprobability sampling using a parametric modeling approach was developed by \citet{kim2021combining}. Nonparametric models were also considered, including nearest-neighborhood estimation \citep{rivers2007} and generalized additive models \citep{yang2021, chen2022nonparametric}. For a comprehensive review of statistical analysis of nonprobability samples, we refer to \citet{bethlehem2010selection, rao2021making, wu2022statistical}.

In this paper, we first establish the relationship between generalized calibration weighting and its dual expression for regression imputation. This dual relationship is critical in identifying the implied regression model and developing model selection for calibration weighting. 
We then develop a unified approach to doubly robust calibration weighting. The calibration weighting is performed in two steps. In the first step, initial weights are constructed using the working propensity score model. These propensity weights are then further adjusted using the debiased calibration method of \citet{kwon2024}. The resulting calibration estimator is doubly robust, meaning that the final estimator is approximately unbiased as long as one of the two models—the propensity score model or the outcome regression model—is correctly specified. The proposed method can be readily extended to construct multiply robust calibration estimation. Asymptotic properties of the proposed two-step calibration estimator are also  investigated. 

The paper is organized as follows. In Section 2, basic setup and the regression weighting method are introduced. In Section 3, the generalized entropy calibration is introduced as a unified framework for calibration weight for handling voluntary survey data. In Section 4, the proposed method is extended to handle two-step calibration. In Section 5, results from a limited simulation study are presented. In Section 6, we present a real data application of the proposed method to data integration problem. 
Some concluding remarks are made in Section 7.


\section{Basic setup}

Consider a finite population with index set $U = \{ 1, \ldots, N \}$.  We consider the situation where certain auxiliary variables are observed throughout the finite population. We use $\bx_i \in \mathbb{R}^p$ and $y_i$ to denote the vector of auxiliary variables and the study variable associated with unit $i$, respectively. 
Assume that we have a non-probability sample $S \subset U$ selected from the finite population and observe $y_i$ in the sample. 


 We are interested in estimating $Y=\sum_{i=1}^N y_i$ from the sample.  
We assume that the sampling mechanism is ignorable in the sense that 
\begin{equation}
 \delta \perp Y \mid \bx ,
 \label{mar}
 \end{equation}
 where $$ \delta_i = \left\{
\begin{array}{ll} 
1 & \mbox{ if } i \in S \\
0 & \mbox{ otherwise.}
\end{array}
\right.
$$ 
 Condition (\ref{mar}) is called \emph{missingness at random}  \citep{rubin1976} in the missing data literature.

  We  consider a linear estimator defined as 
$$ \widehat{Y}_\omega = \sum_{i \in S} \omega_i y_i $$
for some $\omega_i>0$ that does not depend on $y$-values. We impose that the final weight satisfy the calibration constraint: 
\begin{equation} 
 \sum_{i \in S} \omega_i
\mathbf{x}_i = \sum_{i=1}^N \mathbf{x}_i.
\label{calib3}
\end{equation} 
We assume that an intercept is included in $\bx_i$ in the sense that $\bx_i^\top \mathbf{a}= 1$ for some $\mathbf{a}$. 
The calibration constraint is justified under the following regression model 
\begin{equation}
y_i = \bx_i^\top \bbeta + e_i
\label{eq:1}
\end{equation}
where $e_i \sim (0, \sigma^2)$ and $e_i$ is independent of $\bx_i$. Note that,   under (\ref{eq:1}), we have  
\begin{equation}
E_{\xi} \left( 
\widehat{Y}_\omega - Y \mid \delta_1, \ldots, \delta_N \right) =  \sum_{i \in S} \omega_i \bx_i^\top \bbeta - \sum_{i=1}^N  \bx_i^\top  \bbeta ,
\label{mean} 
\end{equation} 
where subscript $\xi$ represents the superpopulation model in (\ref{eq:1}). Thus, as long as the calibration condition in (\ref{calib3}) holds, the linear estimator $\widehat{Y}_{\omega}$ is unbiased under the regression model in (\ref{eq:1}).

Now, to uniquely determine $\omega_i$, note that 
\begin{equation}
V_{\xi} \left( 
\widehat{Y}_\omega - Y \mid \delta_1, \ldots, \delta_N \right) =    \sum_{i \in S} \omega_i^2 \sigma^2 - 2   \sum_{i \in S} \omega_i \sigma^2 + 
\sum_{i =1}^N \sigma^2   .
\label{var}
\end{equation} 
\textcolor{black}{Since an intercept is included in $\bx_i$,  (\ref{calib3}) implies that 
$$  
V_{\xi} \left( 
\widehat{Y}_\omega - Y \mid \delta_1, \ldots, \delta_N \right) =    \sum_{i \in S} \omega_i^2 \sigma^2 - 
\sum_{i =1}^N \sigma^2   . $$
} 
Therefore, minimizing the model variance of $\widehat{Y} _\omega- Y$ under model unbiasedness 
is equivalent to minimizing 
\begin{equation}
Q(w) = \sum_{i \in S} \omega_i^2   
\label{eq:9-30}
\end{equation}
subject to calibration  constraint in (\ref{calib3}).

The resulting calibration  estimator that minimizes (\ref{eq:9-30}) subject to (\ref{calib3})  is then given by 
\begin{equation}
\widehat{Y}_{\rm cal} = \sum_{i \in S} \hat{\omega}_i y_i = \sum_{i=1}^N \bx_i^\top  \hat{\bbeta} 
\label{res3}
\end{equation}
where 
$$ \hat{\omega}_i = \sum_{i=1}^N \bx_i^\top \left( \sum_{i \in S}  \bx_i \bx_i^\top   \right)^{-1}  \bx_i$$
and 
$
\hat{\bbeta}  = \left( \sum_{i \in S}  \bx_i \bx_i^\top    \right)^{-1}\sum_{i \in S}  \bx_i y_i 
.$ 
Expression (\ref{res3})  reveals that the final calibration estimator can be interpreted as the regression prediction  estimator under the regression model (\ref{eq:1}). In other words, the calibration estimator satisfying constraint (\ref{calib3}) uses a regression  model in (\ref{eq:1}) implicitly. See  \cite{fuller2002} for a comprehensive treatment on regression estimation. 


\section{Generalized entropy calibration}

We now use the generalized entropy \citep{gneiting2007strictly} to develop a class of calibration estimators. Let  $G( \nu)$ be a strictly convex and differentiable function in a domain $\mathcal{D} \subset \mathbb{R}$.
We now consider the generalized entropy calibration (GEC) weighting by minimizing 
\begin{equation}
\sum_{i \in S}  G \left( {\omega_i} \right) 
\label{wel}
\end{equation}
    subject to (\ref{calib3}). 
    Once the GEC weight $\hat{\omega}_i$ are obtained, the GEC estimator of $Y=\sum_{i=1}^N y_i$ is given by $\widehat{Y}_{\rm GEC} = \sum_{i \in S} \hat{\omega}_i y_i$. \textcolor{black}{The regression calibration estimator in (\ref{res3}) is a special case of the GEC estimator in (\ref{wel}) with $G(\omega) = \omega^2$. }

    Using the Lagrange multiplier method, we find the minimizer of 
    $$ \mathcal{L} ( \bomega, \blambda) =  \sum_{i \in S} G( \omega_i) - \blambda^\top \left( \sum_{i \in S} \omega_i \bx_i - \sum_{i=1}^N \bx_i \right) $$
    with respect to $\blambda$ and $\bomega$. 
    By setting 
    $\partial  \mathcal{L} / \partial \omega_i= 0$
    and solving for $\omega_i$, we obtain 
    \begin{equation}
     \hat{\omega}_i ( \blambda) = g^{-1} \left( \bx_i^\top \blambda \right) , 
     \label{eq:3-2}
     \end{equation}
    where $g(\omega) = d G( \omega) / d \omega$. \textcolor{black}{The derivation for (\ref{eq:3-2}) is the same as that of \cite{deville1992} using a constant design weight. }

    By plugging $\hat{\omega}_i ( \blambda)$ into $\mathcal{L}$,    
    we can formulate a  dual optimization problem: 
 \begin{equation}
 \hat{\blambda} = \mbox{arg} \min_{\lambda} \left[   \sum_{i \in S}  \rho \left( \bx_i^\top \blambda \right) - \sum_{i=1}^N \bx_i^\top \blambda  
 \right] ,  
 \label{dual}
 \end{equation}
 where 
 $ \rho \left( \nu \right) $
 is the \emph{convex conjugate function}  of $G$, which is defined by 
 $$ \rho \left( \nu \right) = \nu \cdot g^{-1} ( \nu) - G \{ g^{-1} ( \nu) \}.$$
 Note that $\rho( \nu)$ satisfies $\rho^{(1)} ( \nu) = g^{-1} ( \nu)$, where $\rho^{(1)} (\nu)  = d \rho(\nu)/ d \nu$.  
  See Table \ref{tab:15-1} for examples of the generalized entropy functions and their convex conjugate functions. 

\begin{table}[ht]
\centering
\begin{tabular}{ccccc}
\hline
Generalized Entropy                                             & $G(\omega)$                 & $\rho(\nu)$     & $\rho^{(1)} ( \nu)$      \\ \hline
Squared loss                                       & $\omega^2/2$                  & $\nu^2/2$              & $\nu$                   \\  
Kullback-Leibler & $\omega \log (\omega)$ & $\exp ( \nu -1)$ &  $\exp ( \nu -1 )$ \\ 
Shifted KL & $(\omega-1) \{ \log (\omega-1)- 1\}$ & $
\nu + \exp ( \nu )$ &  $1+ \exp ( \nu )$ \\
Empirical likelihood & $- \log ( \omega)$ & $-1- \log ( -\nu), \nu<0$  & $-1/ \nu$ \\
Hellinger distance & $( \sqrt{\omega} -1 )^2$ & $\nu/(1-\nu), \nu <1$ & $(1-\nu)^{-2}$ \\
R\'enyi entropy($\alpha \neq 0, -1$) & $ \frac{1}{\alpha+1} \omega^{\alpha + 1}$   & $ \frac{\alpha}{\alpha+1} \nu^{\frac{\alpha+1}{\alpha}}, \nu>0$ &  $\nu^{1/\alpha}$\\
\hline
\end{tabular}
\caption{Examples of generalized entropies, $G(\omega)$, and the corresponding convex conjugate functions, and their derivatives } 
\label{tab:15-1}
\end{table}

 Now, by (\ref{dual}), we have 
\begin{equation}
\sum_{i \in S}  \rho^{(1)} ( \bx_i^\top \hat{\blambda} ) \bx_i =  \sum_{i=1}^N \bx_i,
\label{cal2}
\end{equation}   
which is the calibraton equation in (\ref{calib3}).   
 Once $\hat{\blambda}$ is obtained from (\ref{dual}), we can obtain 
 $$ \widehat{Y}_{\rm GEC} = \sum_{i \in S}  \rho^{(1)} \left( \bx_i^{\top} \hat{\blambda} \right) y_i $$
as the proposed calibration estimator. The following theorem presents the asymptotic properties of the calibration estimator using generalized entropy calibration.

\begin{theorem}
\label{thm:1}
Let $\hat{\omega}_i$ be obtained by minimizing (\ref{wel}) subject to (\ref{calib3}).  Under some regularity conditions, the resulting calibration estimator 
$ \widehat{Y}_{\rm GEC} = \sum_{i \in S}  \hat{\omega}_i y_i $ 
satisfies 
\begin{equation}
 \widehat{Y}_{\rm GEC}  = \sum_{i=1}^N \eta_i + o_p \left( n^{-1/2} N \right) 
\label{result}
\end{equation}
where 
$$ \eta_i = \bx_i^\top \bgamma^*   + \delta_i \rho^{(1)} \left( \bx_i^\top \blambda^* \right) \left( y_i - \bx_i^{\top} \bgamma^* \right), $$
$\blambda^* = p \lim \hat{\blambda}$ , $\bgamma^* = p \lim \hat{\bgamma}$, and 
\begin{equation}
\hat{\bgamma} = \left\{ \sum_{i \in S}  \hat{q}_i  \bx_i \bx_i^{\top} \right\}^{-1} \sum_{i \in S} \hat{q}_i \bx_i y_i 
\label{gammahat}
\end{equation}
with  $\hat{q}_i = \rho^{(2)}(\bx_i^\top \hat{\blambda})$ and $\rho^{(2)} (\nu) = d^2  \rho ( \nu) / d \nu^2 $. 
\end{theorem}

Due to covariate balancing, the calibration estimator $\widehat{Y}_{\rm GEC}$ can also be written as
\begin{equation}
 \widehat{Y}_{\rm GEC}  = \sum_{i=1}^N \bx_i^{\top} \hat{\bgamma} + \sum_{i \in S} \hat{\omega}_i \left( y_i - \bx_i^{\top} \hat{\bgamma}  \right),  
\label{result1}
\end{equation}
where $\hat{\bgamma}$ is defined in (\ref{gammahat}). Note that the calibration weight can be expressed as $\hat{\omega}_i = \rho^{(1)}(\bx_i^\top \hat{\blambda})$, and that the linearization result in (\ref{result}) does not rely on any model assumptions. From the expression in (\ref{result1}), the proposed calibration estimator consists of two components: the first is a prediction term obtained by performing a weighted regression of $y_i$ on $\bx_i$ using weights $\hat{q}_i$, and the second is an adjustment term that provides protection against model misspecification.

This second term may be interpreted as a debiasing term, particularly when $\hat{\omega}_i^{-1}$ serves as an estimate of the first-order inclusion probability for the sample $S$. The regression coefficients $\hat{\bgamma}$ in the linearization of the proposed GEC estimator are computed using weights $\hat{q}_i = \rho^{(2)}(\bx_i^\top \hat{\blambda})$, which is derived by differentiating the calibration weight $\hat{\omega}_i = \rho^{(1)}(\bx_i^\top \hat{\blambda})$ with respect to $\blambda$. Because the first-order effect of $\hat{\blambda}$ on $\hat{\omega}_i$ is absorbed into the estimation of $\hat{\bgamma}$ in (\ref{gammahat}), the use of $\hat{q}_i$ in the computation of $\hat{\bgamma}$ is theoretically well-justified.
 
Therefore, the  consistency of $\widehat{Y}_{\rm GEC}$ can be established under one of the two model assumptions.
\begin{enumerate}
\item Outcome regression (OR) model in (\ref{eq:1}). 
\item Propensity score (PS) model given by 
\begin{equation}
 P \left( \delta_i =1 \mid \bx_i \right) = \left\{ \rho^{(1)} \left( \bx_i^\top \bphi_0 \right) \right\}^{-1} 
\label{ps}
\end{equation}
for some $\bphi_0$.
\end{enumerate}

 If the outcome regression model in (\ref{eq:1}) holds, then $\bgamma^*= \bbeta$ and 
$$ \eta_i = \bx_i^{\top}\bbeta+   \delta_i \omega_i^* \left( y_i - \bx_i^\top \bbeta \right) , $$
where $\omega_i^* =\rho^{(1)} \left( \bx_i^\top \blambda^* \right) $.  Thus, we can write 
$$\widehat{Y}_{\rm GEC} - Y  \cong \sum_{i=1}^N \eta_i - \sum_{i=1}^N y_i = 
\sum_{i=1}^N \left( \delta_i \omega_i^* - 1 \right) \left(y_i - \bx_i^{\top} \bbeta \right),$$
%
%
which is unbiased to zero under the outcome regression model in (\ref{eq:1}) and the missing at random (MAR) assumption in (\ref{mar}). Under those conditions, 
$\widehat{Y}_{\rm GEC}$ is asymptotically model-unbiased for $Y=\sum_{i=1}^N y_i$ with 
\begin{eqnarray}
    N^{-1} V \left( \widehat{Y}_{\rm GEC} - Y  \right) & \cong & 
    \frac{1}{N} \sum_{i=1}^N 
E \left( \delta_i \omega_i^* - 1 \right)^2 \sigma^2 \notag  \\
& \cong & 
\frac{1}{N} \sum_{i \in S} （\omega_i^{* 2} - \omega_i^*） \sigma^2
\ \cong \ 
\frac{1}{N} \sum_{i \in S} \omega_i^{* 2} \sigma^2 - \sigma^2, \label{vargec2}
\end{eqnarray}
where the variance is with respect to $(\delta_i, \bx_i, y_i)$ in the superpopulation of (\ref{eq:1}), and the last two approximations are due to the covariate balancing constraint $\sum_{i \in S} \hat{\omega}_i = N$ in (\ref{cal2}) if the constant 1 is included in the covariates $\bx_i$.

The variance expression in (\ref{vargec2}) warrants further discussion. Without loss of generality, suppose that the outcome model in (\ref{eq:1}) can be re-expressed as
\[
y_i = \bx_{i1}^\top \bm{\beta}_1 + \bx_{i2}^\top \bm{\beta}_2 + e_i,
\]
where $\bm{\beta}_2 = \mathbf{0}$. In this setting, the covariates $\bx_{i2}$ do not contribute to the prediction of $y_i$. Including such irrelevant covariates in the calibration constraints may inflate the term $(\omega_i^*)^2$ without reducing the model variance, especially for the square entropy $G(\omega) = \omega^2$. Consequently, the variance in (\ref{vargec2}) is minimized when only the relevant covariates $\bx_{i1}$ are used for calibration.

This observation highlights a potential drawback of over-calibration: including unnecessary calibration variables can increase estimator variance without improving bias reduction. This conclusion aligns with the empirical findings of \citet{brookhart2006variable} and \citet{shortreed2017}.

Now, under the PS model in (\ref{ps}), we can obtain $\blambda^*= \bphi_0$ and 
$$ \eta_i = \bx_i^\top \bgamma^*   + \frac{\delta_i}{\pi_i}   \left( y_i - \bx_i^{\top} \bgamma^* \right)$$
where $\pi_i = P( \delta_i =1 \mid \bx_i)$. 
Thus, we can establish 
$E \left( \eta_i \mid \bx_i, y_i \right) = y_i$, where the conditional distribution is with respect to the probability law in $[ \delta \mid \bx, y]$. Under the PS model in (\ref{ps}), ignoring the smaller order terms, we have 
\begin{equation}
N^{-1} V \left( \widehat{Y}_{\rm GEC} - Y \right) \cong  E\left\{ \frac{1}{N} \sum_{i=1}^N ( \pi_i^{-1} -1 ) ( y_i - \bx_i^\top \bgamma^* )^2 \right\},
\label{result3}
\end{equation}
where the variance is with respect to $\delta_i$ given $(\bx_i, y_i)$.
Note that  (\ref{result3}) is equal to the Godambe-Joshi lower bound of the anticipated variance under the model in (\ref{eq:1}).

Therefore, the proposed GEC estimator is locally efficient and doubly robust, where the double robustness means that its consistency is justified as long as one of the two models, OR model in (\ref{eq:1}) and PS model in (\ref{ps}), is correctly specified. Doubly robust estimation is very popular in the missing data literature \citep{bang05, cao09, 
kimhaziza12}. 

From the expressions in (\ref{vargec2}) and (\ref{result3}), we can estimate the variance $V( \widehat{Y}_{\rm GEC} - Y)$ of the proposed GEC estimator by
%
%
\begin{equation} \hat{V} = \sum_{i \in S} \hat{\omega}_i ( \hat{\omega}_i -1 )  ( y_i - \bx_i^\top \hat{\bgamma})^2, \label{varest2}
\end{equation}
which is similar to the variance estimator in \cite{kwon2024} under Poisson sampling.
Note that the variance estimator in (\ref{varest2}) is doubly robust, in the sense that it remains consistent if either the outcome model or the propensity score model is correctly specified. 
Its implementation only needs the data in the sample and the summary statistics $\sum_{i=1}^N \mathbf{x}_i$ for covariate balancing, which doesn't require access to individual-level covariate information $\bx_i$ for all units in the population. 

\begin{remark}
 The information projection approach of \citet{wang2024} is equivalent to using 
     \begin{equation}
      G( u) =(u-1) \{  \log (u-1) -1 \}.
      \label{eq:15}
      \end{equation}
The convex conjugate  function of $G(u)$ is 
\begin{equation}
 \rho ( \nu) = \nu + \exp (  \nu) 
 \label{tan}
 \end{equation}
which leads to the calibrated maximum likelihood method of \citet{tan2020}. 
\end{remark}

{In practice, avoiding extremely large  weights is an important practical problem \citep{chen2002, ma2020}. Thus, we wish to have the final weights  bounded. That is, we wish to achieve 
$$ \hat{\omega}_i ( \blambda ) = \rho^{(1)} ( \bx_i^\top \blambda ) \le  M $$
for some $M$ which is the upper bound of the final weight.  To achieve the goal, one way is to  use a Huber's loss function 
\begin{equation}
\rho(\nu) = \begin{cases}
    \dfrac{1}{2} \nu^2 & \abs{\nu} \leq M \\
    M \del{\abs{\nu} -\dfrac{1}{2} M} & \abs{\nu} > M 
\end{cases}
\label{huber}
\end{equation}
and obtain $G$ corresponding to $\rho$:
\begin{equation}
    G(\omega) = \begin{cases}
    \dfrac{\omega^2}{2} & \abs{\omega} \leq M \\
    \infty & \abs{\omega} > M 
\end{cases} \label{123}
\end{equation}
or use $G(\omega) = \omega^2 /2$ and add extra soft calibration constraint $\abs{\omega} \leq M$ which is equivalent to \eqref{123}. 

\textcolor{black}{
To determine $M$, 
we can treat $M$ as a tuning parameter in the weighted regression and choose an optimal $M$  in a data-driven way. Note that the final weights using (\ref{123}) can be expressed as 
\begin{equation}
 \hat{\omega}_i^*( \blambda, M ) = \min \left\{\rho^{(1)} ( \bx_i^\top \blambda ) 
 , M \right\},
\label{trim}
\end{equation}
which is essentially weight trimming. Now,  
similiar to \eqref{result1}, we can approximate the final calibration estimator as a GREG  estimator  
$$ \sum_{i \in S} \hat{\omega}_i^* ( \hat{\blambda}, M ) y_i \cong 
  \sum_{i=1}^N \bx_i^{\top} \hat{\bbeta}_M+ \sum_{i \in S} \hat{\omega}_i^*( \blambda, M ) \left( y_i - \bx_i^{\top} \hat{\bbeta}_M \right), $$
where 
\begin{equation}
 \hat{\bbeta}_M = \left\{ \sum_{i \in S} \hat{q}_i^*( \hat{\blambda}, M ) \bx_i \bx_i^{\top} \right\}^{-1}\sum_{i \in S} \hat{q}_i^*( \hat{\blambda}, M ) \bx_i y_i 
 \label{beta}
 \end{equation}
and 
$$\hat{q}_i^*( \hat{\blambda}, M ) = \left\{ 
\begin{array}{ll}
\rho^{(2)} ( \bx_i^\top \hat{\blambda} ) 
 & \mbox{ if }  \rho^{(1)} ( \bx_i^\top \hat{\blambda} ) \le M  \\
 0 & \mbox{otherwise}. 
\end{array} 
\right.
$$
Therefore, standard cross-validation method can be used to determine the optimal $M$ that minimizes the mean squared prediction errors using $\hat{y}_i = \bx_i^{\top} \hat{\bbeta}_M$. }


\color{black}

\begin{remark}\label{rm:1}
Our proposed calibration method is similar to the nonparametric calibration weighting method of 
\cite{chan2016} in the context of causal inference. In \cite{chan2016}, the balancing  function used for calibration weighting is the basis function for sieve function approximation. In causal inference, the study variable $Y$ of interest is often univariate. Thus, nonparametric regression approach is a sensible  choice. However, in survey sampling, we collect many study variables and developing nonparametric regression for multivariate $Y$ is quite challenging. Thus, we use parametric regression model   as a working model for calibration. 
\end{remark}

\section{Two-step calibration}

Since the proposed GEC estimator is justified under two different models,  we can incorporate the two working models  separately to develop a two-step calibration, where the PS model is used in the first step and the OR model is used in the second step. By constructing the final weights using the OR model only, we may reduce the variability of the final  weights and achieve statistical efficiency.

To explain the proposed method, we consider the following working models:
\begin{enumerate}
\item \textbf{The propensity score model:}
    \begin{equation}
     P ( \delta_i =1 \mid \bx_i ) = \pi( \bx_{1i}^{\top} \bphi_0 ),
     \label{ps2}
     \end{equation}
    where $\bphi_0$ is a parameter vector, $\pi(\nu) \in (0,1]$ is a known monotone function, and $\bx_{1i}$ is a subvector of $\bx_i$.
    
\item \textbf{The outcome regression model:}
    \begin{equation}
     y_i = \bx_{2i}^{\top} \bbeta + e_i,
     \label{or2}
     \end{equation}
    where $e_i \mid \bx_i \sim (0, \sigma^2)$ and $\bx_{2i}$ is a subvector of $\bx_i$.
\end{enumerate}

Note that the covariates used in each model are subsets of the full covariate vector. The subvectors $\bx_{1i}$ and $\bx_{2i}$
may overlap. Under the reduced outcome regression model in (\ref{or2}), the MAR assumption implies that $Y$ and $\delta$ are conditionally independent given $\bx_{1i}$. See Lemma 3 of \cite{wang2024} for details.

Our first objective is to construct a calibration weighting method in which the calibration equation
\begin{equation}
\sum_{i \in S} \frac{1}{\pi \left( \bx_{1i}^\top \bphi \right) } \bx_{1i} = \sum_{i=1}^N \bx_{1i}
\label{ps-cal}
\end{equation}
can be interpreted as the solution to a dual optimization problem, analogous to (\ref{dual}). Specifically, we aim to identify a convex conjugate function $\rho$ such that the calibration equation (\ref{ps-cal}) arises as the first-order condition for minimizing the following objective:
\begin{equation}
F (\bphi) = \sum_{i \in S} \rho \left(\bx_{1i}^{\top} \bphi \right)
-  \sum_{i=1}^N \bx_{1i}^{\top} \bphi.
\label{dual-2}
\end{equation}
To ensure this equivalence, the function $\rho$ must satisfy the condition
\begin{equation}
\frac{d}{d \nu} \rho (\nu) = \left\{ \pi ( \nu) \right\}^{-1}.
\label{adjust}
\end{equation}
In addition, for $\rho$ to be convex, the function $\{\pi( \nu)\}^{-1}$ must be monotone increasing in $\nu$.

Since the objective function $F( \bphi)$ in (\ref{dual-2}) induces the calibration equation in (\ref{ps-cal}), we refer to it as the \emph{calibration generating function} (CGF). Notably, the calibrated likelihood function proposed by \cite{tan2020} is a special case of our CGF when $\pi( \bx_{1}^\top \bphi)$ follows the logistic regression model.

Once the function $\rho$ is specified, we can derive its convex conjugate to construct a primal optimization formulation for the propensity score (PS) weighting under model (\ref{ps2}). The convex conjugate function of $\rho$, which satisfies the condition in (\ref{adjust}), is given by
\begin{equation}
G_1(u) = u \, \pi^{-1}(1/u) - \rho\{ \pi^{-1}(1/u) \}.
\label{dual2a}
\end{equation}

Equation (\ref{dual2a}) follows from the definition of the convex conjugate of $\rho$, under the derivative constraint in (\ref{adjust}). For example, suppose
\[
\pi(\nu) = \{ 1 + \exp(\nu) \}^{-1},
\]
in which case $\pi(\nu)$ is monotone decreasing in $\nu$, and the corresponding function $\rho$ becomes
\[
\rho(\nu) = \nu + \exp(\nu).
\]
The inverse function of $\pi$ evaluated at $1/u$ is given by
\[
\pi^{-1}(1/u) = \log(u - 1).
\]
Substituting into (\ref{dual2a}) yields
\begin{align*}
G_1(u) &= u \, \pi^{-1}(1/u) - \rho\{ \pi^{-1}(1/u) \} \\
       &= u \log(u - 1) - \left[ \log(u - 1) + (u - 1) \right] \\
       &= (u - 1) \log(u - 1) - (u - 1),
\end{align*}
which corresponds to the entropy function $G_1(\omega)$ defined in equation (\ref{eq:15}). 
Additional examples of common inverse link functions $\pi(\nu)$ and their corresponding entropy functions are provided in Table~\ref{tablink}.

\begin{table}[!ht]
\centering
\begin{tabular}{@{}ccc@{}}
\toprule
Link                  & $\pi(-\nu)$                                   & $G(\omega)$                              \\ \midrule
Logistic              & $\{ 1+ \exp (-\nu) \}^{-1}$                   & $(\omega - 1)\cbr{\log(\omega - 1) - 1}$ \\
Log                   &  $\exp(\nu)$  &   $\omega \log \omega -\omega$ \\
Identity                   &  $\nu$  &   $-\log \omega$ \\
Probit                & $\Phi(\nu)$                                 &  $-\bigints_{-\infty }^{\omega} \Phi^{-1}\del{\dfrac{1}{t}}dt$  \\
\begin{tabular}[c]{@{}c@{}}Complementary \vspace{-1em}\\  log-log\end{tabular} & $\cbr{1 - \exp \del{-\exp(\nu)}}$           &  $-\bigints_{-\infty }^{\omega} \log \del{\log \dfrac{t}{t-1}}dt$  \\
Cauchy                & $\dfrac{1}{2} + \dfrac{1}{\pi}\arctan(\nu)$ &  $-\bigints_{-\infty }^{\omega} 
\cot \del{\dfrac{\pi}{t}} dt$ \\
Power($\alpha \neq -1$)          &  $\nu^{-1/\alpha}$  &   $\dfrac{1}{\alpha+1}\omega^{\alpha + 1}$                                       \\ \bottomrule
\end{tabular}
\caption{The inverse link functions $\pi(\cdot)$ for modeling the propensity scores and their induced entropies $G(\cdot)$.}
\label{tablink}
\end{table}

Therefore, we adopt the following two-step approach to GEC calibration. 
\begin{description}
\item{[Step 1]} Given $\pi(\cdot)$ as specified in (\ref{ps2}), find a function $\rho(\nu)$ such that condition (\ref{adjust}) holds. Using (\ref{dual2}), we determine the GEC weights by solving the following optimization problem:
\[
\{\hat{\omega}_{1i} \} = \argmin_{\{\omega_i\}} \sum_{i \in S} G_1(\omega_i) \quad \text{subject to} \quad
\sum_{i \in S} \omega_i \bx_{1i} = \sum_{i=1}^N \bx_{1i}.
\]
If $\pi( \nu)$ is monotone increasing, consider $\tilde{\pi}(\nu)=\pi(- \nu)$ and apply (\ref{adjust}) and (\ref{dual2a}) using  $\tilde{\pi} ( \nu)$ in place of $\pi( \nu)$.

\item{[Step 2]} Using the weights $\hat{\omega}_{1i}$ obtained from Step 1, compute the second-step GEC weights by solving:
\[
\{\hat{\omega}_{2i}\} = \argmin_{\{\omega_i\}} \sum_{i \in S} G_2(\omega_i)
\]
subject to the calibration constraints:
\begin{align}
\sum_{i \in S} \omega_i \bx_{2i} &= \sum_{i =1}^N \bx_{2i}, \label{cal-1} \\
\sum_{i \in S} \omega_i g_2(\hat{\omega}_{1i}) &= \sum_{i =1}^N g_2(\hat{\omega}_{1i}), \label{cal-2}
\end{align}
where $g_2(\omega) = dG_2(\omega)/d\omega$.
\end{description}

\color{black}

The purpose of the first-step calibration is to compute the initial propensity score (PS) weights that reflect the PS model specified in (\ref{ps2}). The second-step calibration serves as the primary calibration step, incorporating both the outcome regression model in (\ref{or2}) and the PS model simultaneously. The additional calibration constraint in (\ref{cal-2}) is introduced to appropriately integrate the initial PS weights into the second-step calibration, thereby ensuring consistency under the PS model. This constraint, given in (\ref{cal-2}), is referred to as the \emph{debiasing calibration constraint} \citep{kwon2024}.
Note that the debiasing constraint in Step 2 depends on an initial estimate of the propensity score. So, it cannot be applied in Step 1.
 

Let $\rho_k(\nu)$ denote the convex conjugate function of $G_k(u)$ for $k = 1, 2$. The final weight $\hat{\omega}_{2i}$, obtained from the two-step calibration procedure, can be expressed as
\[
\hat{\omega}_{2i} = \rho_2^{(1)} \left( \hat{\bz}_i^{\top} \hat{\blambda} \right),
\]
where $\hat{\bz}_i^{\top} = \left( \bx_{2i}^{\top}, \hat{g}_{2i} \right)$, and
\[
\hat{g}_{2i} = g_2 \left( \hat{\omega}_{1i} \right) = g_2 \left\{ \rho_1^{(1)} \left( \bx_{1i}^{\top} \hat{\bphi} \right) \right\}.
\]
The vector $\hat{\blambda}$ is obtained by solving the following optimization problem:
\begin{equation}
\hat{\blambda} = \arg\min_{\blambda} \left[ \sum_{i \in S} \rho_2 \left( \hat{\bz}_i^\top \blambda \right) - \sum_{i=1}^N \hat{\bz}_i^\top \blambda \right].
\label{dual2}
\end{equation}

We now discuss the asymptotic properties of the two-step GEC estimator. Let $$\widehat{Y}_{\rm GEC2} = \sum_{i \in S} \hat{\omega}_{2i} y_i$$ denote the resulting estimator based on the final weights. By applying Theorem 1 and treating $\hat{\bphi}$ as fixed (i.e., ignoring its estimation uncertainty), we obtain the following linearization:
\begin{equation}
\widehat{Y}_{\rm GEC2}
= \sum_{i=1}^N \eta_{2i} + o_p\left( n^{-1/2} N \right),
\label{eq:12-res1}
\end{equation}
where
\begin{equation*}
\eta_{2i} = \hat{\bz}_i^\top \bgamma_2^* + \delta_i \, \rho_2^{(1)} \left( \hat{\bz}_i^\top \blambda^* \right) \left( y_i - \hat{\bz}_i^\top \bgamma_2^* \right).
\end{equation*}
Here, $\bz_i^\top = \left( \bx_{2i}^\top, \hat{g}_{2i} \right)$, where $\hat{g}_{2i} = g_2(\hat{\omega}_{1i})$, and $\blambda^* = p \lim \hat{\blambda}$ and $\bgamma_2^*= \hat{\bgamma}_2$ where 
\begin{equation}
\hat{\bgamma}_2 = \left( \sum_{i \in S} \rho_2^{(2)}(\bz_i^\top \hat{\blambda}) \, \bz_i \bz_i^\top \right)^{-1} \sum_{i \in S} \rho_2^{(2)}(\bz_i^\top \hat{\blambda}) \, \bz_i y_i.
\label{gamma2}
\end{equation}

Under the outcome regression model specified in (\ref{or2}), we have
\[
\bz_i^\top \bgamma_2^* = \bx_{2i}^\top \bbeta,
\]
which implies that
\[
\eta_{2i} = \bx_{2i}^\top \bbeta + \delta_i \, \omega_{2i}^* \left( y_i - \bx_{2i}^\top \bbeta \right),
\]
where $\omega_{2i}^* = \rho_2^{(1)} \left( \bz_i^\top \blambda^* \right)$. Therefore, under the outcome regression model (\ref{or2}), the two-step GEC estimator $\widehat{Y}_{\rm GEC2}$ is approximately unbiased, and its approximate variance is given by
\[
N^{-1} V \left( \widehat{Y}_{\rm GEC2} - Y \right)
\cong \frac{1}{N} \sum_{i \in S} \omega_{2i}^{* 2} \sigma^2  - \sigma^2.
\] 
Because only the true covariates $\bx_2$ are directly used in the calibration process, the two-step GEC estimator achieves greater efficiency than the original GEC estimator when the reduced outcome model (\ref{or2}) is correctly specified.

On the other hand, under the PS model in (\ref{ps2}),  we can establish the following theorem.  
\begin{theorem}
Assume that the PS model in (\ref{ps2}) holds. 
Under some regularity conditions,  the two-step calibration estimator 
$ \widehat{Y}_{\rm GEC2} = \sum_{i \in S}  \hat{\omega}_{2i} y_i $ 
satisfies the following asymptotic linearization:
\begin{equation}
\widehat{Y}_{\rm GEC2} = \sum_{i=1}^N \eta_{2i} + o_p\left(n^{-1/2} N\right),
\label{result2}
\end{equation}
where
\[
\eta_{2i} = \bx_{1i}^\top \bgamma_1^* + \bz_i^\top \bgamma_2^* + \frac{\delta_i}{\pi(\bx_{1i}^\top \bphi_0)} \left( y_i - \bx_{1i}^\top \bgamma_1^* - \bz_i^\top \bgamma_2^* \right),
\]
with $\bz_i^\top = \left( \bx_{2i}^\top, g_{2i} \right)$, $g_{2i} = g_2 \left\{ \rho_1^{(1)}(\bx_{1i}^\top \bphi_0) \right\}$, and $\bphi_0$ denoting the true value of the PS model parameter. 
The vectors $\bgamma_1^*$ and $\bgamma_2^*$ are the probability limits of the solution $(\bgamma_1, \bgamma_2)$ to the following system of normal equations:
\[
\begin{pmatrix}
\sum_{i \in S} \hat{q}_i \bx_{1i} \bx_{1i}^\top & \sum_{i \in S} \hat{q}_i \bx_{1i} \hat{\bz}_i^\top \\
\sum_{i \in S} \hat{q}_i \hat{\bz}_i \bx_{1i}^\top & \sum_{i \in S} \hat{q}_i \hat{\bz}_i \hat{\bz}_i^\top
\end{pmatrix}
\begin{pmatrix}
\bgamma_1 \\
\bgamma_2
\end{pmatrix}
=
\begin{pmatrix}
\sum_{i \in S} \hat{q}_i \bx_{1i} y_i \\
\sum_{i \in S} \hat{q}_i \hat{\bz}_i y_i
\end{pmatrix},
\]
where $\hat{\bz}_i = \left( \bx_{2i}^\top, \hat{g}_{2i} \right)^\top$ and $\hat{q}_i = \rho_2^{(2)}(\hat{\bz}_i^\top \hat{\blambda})$.
\end{theorem}

Theorem 2 establishes the consistency of the GEC estimator under the PS model. 
A consistent variance estimator can be constructed similarly to (\ref{varest2}) using $\hat{\omega}_{2i}$ and $\bx_{1i}^\top \hat{\bgamma}_1 + \hat{\bz}_i^\top \hat{\bgamma}_2$ instead of $\hat{\omega}_i$ and $\bx_i^\top \hat{\bgamma}$. 

\begin{remark}
In Step 1, covariate balancing is enforced for $\bx_1$ through a calibration condition. Alternatively, one could consider a maximum likelihood estimation (MLE) approach. Under MLE, the score equation for $\bphi$ takes the form
\[
\sum_{i \in S} \mathbf{h}_{1i}(\bphi) = \sum_{i=1}^N \pi\left( \bx_{1i}^{\top} \bphi \right) \mathbf{h}_{1i}(\bphi),
\]
where $\mathbf{h}_{1i}(\bphi) =  \partial  \mbox{logit} \left\{ \pi\left( \bx_{1i}^{\top} \bphi \right) \right\}/ \partial \bphi$. Note that this score equation requires access to the full set of individual covariate values $\bx_{1i}$ for the entire finite population. In contrast, the calibration condition in (\ref{calib3}) requires only the population total $\sum_{i=1}^N \bx_{1i}$ to be known. In practice, due to confidentiality and privacy constraints, analysts rarely have access to unit-level auxiliary information across the full population. Therefore, in the analysis of voluntary or non-probability survey data, the calibration-based approach  is typically more practical and attractive than the MLE-based alternative.
\end{remark}

\section{Simulation study}

To evaluate the performance of the GEC estimators, we used the 2021 National Health Information Database (NHID), a public dataset created by the National Health Insurance Service (NHIS) that contains health status and demographic information for 1,000,000 randomly selected individuals in the Republic of Korea. From this database, we generated a pseudo-population of size $N = 100,000$ through random sampling. Following standard panel survey methodology, which employs stratified sampling based on region, age group, and sex, we generated  samples of size $n=2,297$ from $17(\mbox{Region}) \times 14(\mbox{Age Group}) \times 2(\mbox{Sex}) = 476$ strata. The sample size for each stratum is $n_h = 5$ if $N_h > 15$ and $n_h = \left\lfloor {N_h / 3} \right\rfloor$ if $N_h \leq 15$, where $N_h$ is the stratum size. This sampling process was repeated 500 times. We analyzed two study variables: a continuous variable $y$ representing hemoglobin levels (g/dL) and a discrete variable $y$ indicating whether an oral examination was conducted.

We evaluated three propensity score (PS) models for estimating the population mean: (PS1) oracle propensity scores, which use the true sample inclusion probabilities; (PS2) equal propensity scores, calculated as the ratio of sample size to population size; and (PS3) propensity scores estimated using a logistic regression model based on the main effects of demographic covariates (region, age group, and sex). In PS3, the first step weights were constructed using calibrated maximum likelihood estimation (CMLE) following \citet{tan2020}, as detailed in Remark 1. Note that both PS2 and PS3 are misspecified. In particular, PS3 is misspecified because it excludes interaction terms among the covariates, which are necessary for correctly modeling the strata formation.


\begin{table}
\centering
\begin{tabular}{@{}lrrrrrrrrrrr@{}}
\toprule
                                                                                           & \multicolumn{1}{l}{} & \multicolumn{5}{c}{\begin{tabular}[c]{@{}c@{}}Hemoglobin(g/dL)\\ (continuous $y$)\end{tabular}} & \multicolumn{5}{c}{\begin{tabular}[c]{@{}c@{}}Oral Exam(0 or 1)\\ (discrete $y$)\end{tabular}} \\ \midrule
                                                                                           &                      & Bias               & SE               & RMSE              & RB               & CR(\%)            & Bias              & SE               & RMSE             & RB                & CR(\%)            \\ \midrule
\multirow{5}{*}{\begin{tabular}[c]{@{}l@{}}PS1\\ (Oracle)\end{tabular}}                    & IPW                  & -0.29              & 3.68             & 3.69              & 0.08             & 96.4             & -0.05             & 1.68             & 1.68             & -0.08             & 94.6             \\
                                                                                           & EL                   & -0.23              & 3.72             & 3.73              & -0.01            & 94.6             & -0.03             & 1.71             & 1.71             & -0.16             & 92.2             \\
                                                                                           & HD                   & -0.24              & 3.69             & 3.69              & -0.00            & 95.0             & -0.02             & 1.68             & 1.68             & -0.15             & 93.2             \\
                                                                                           & SL                   & -0.33              & 3.65             & 3.67              & 0.08             & 96.0             & -0.03             & 1.67             & 1.67             & -0.09             & 93.4             \\
                                                                                           & SKL                  & -0.23              & 3.73             & 3.73              & -0.01            & 94.6             & -0.03             & 1.71             & 1.71             & -0.16             & 92.4             \\ \hdashline
\multirow{5}{*}{\begin{tabular}[c]{@{}l@{}} PS2\\ (One-step\\ GEC)\end{tabular}} & IPW                  & -21.11             & 2.33             & 21.23             & 0.19             & 0.0              & -0.89             & 0.90             & 1.27             & 0.14              & 87.4             \\
                                                                                           & EL                   & -3.58              & 2.52             & 4.38              & 0.11             & 74.0             & 1.76              & 1.06             & 2.06             & 0.07              & 66.0             \\
                                                                                           & HD                   & -3.58              & 2.51             & 4.38              & 0.11             & 73.6             & 1.76              & 1.06             & 2.06             & 0.07              & 66.2             \\
                                                                                           & SL                   & -3.72              & 2.51             & 4.49              & 0.12             & 72.0             & 1.76              & 1.05             & 2.05             & 0.08              & 66.0             \\
                                                                                           & SKL                  & -3.58              & 2.52             & 4.38              & 0.11             & 74.0             & 1.76              & 1.06             & 2.06             & 0.07              & 66.0             \\ \hdashline
\multirow{5}{*}{\begin{tabular}[c]{@{}l@{}} PS3\\ (Two-step\\ GEC)\end{tabular}} & IPW                  & -1.28              & 3.69             & 3.91              & 0.07             & 94.6             & -0.18             & 1.65             & 1.65             & -0.08             & 94.6             \\
                                                                                           & EL                   & -1.10              & 3.65             & 3.81              & -0.01            & 94.6             & -0.12             & 1.65             & 1.65             & -0.16             & 92.8             \\
                                                                                           & HD                   & -1.05              & 3.67             & 3.81              & -0.01            & 94.6             & -0.15             & 1.65             & 1.65             & -0.15             & 93.0             \\
                                                                                           & SL                   & -1.05              & 3.67             & 3.82              & 0.07             & 95.0             & -0.19             & 1.65             & 1.66             & -0.09             & 94.4             \\
                                                                                           & SKL                  & -1.11              & 3.64             & 3.81              & -0.01            & 94.4             & -0.12             & 1.64             & 1.65             & -0.16             & 92.8             \\ \bottomrule
\end{tabular}
\caption{Bias($\times 10^2$), standard error(SE, $\times 10^2$), and root mean-squared error(RMSE, $\times 10^2$) of the point estimates, relative bias(RB) of the variance estimates, and the coverage rate(CR, $\times 10^2$) of the 95\% confidence interval.}
\label{table_sim}
\end{table}

Using three different propensity scores, we evaluated five estimators: (1) an inverse probability weighted (IPW) estimator without a calibration constraint, and four GEC estimators based on different entropies: (2) empirical likelihood (EL), (3) Hellinger distance (HD), (4) squared loss (SL), and (5) shifted Kullback-Leibler (SKL). The IPW estimator using PS2 is equivalent to the unweighted sample mean. The GEC estimators based on PS2 correspond to the one-step GEC estimators discussed in Section 3. In contrast, the GEC estimators based on PS3 are the two-step GEC estimators described in Section 4, where SKL entropy is used in the first step. For the GEC estimators, we included ten auxiliary variables for calibration: age group, sex, height, weight, waist, alcohol consumption (binary indicator), systolic blood pressure, diastolic blood pressure, fasting blood sugar, and creatinine. For variance estimation of the GEC estimators, we used \eqref{varest2}.

Table \ref{table_sim} summarizes the performance of the estimators in terms of bias, standard error (SE), and root mean squared error (RMSE) of the point estimates, along with the relative bias (RB) of the variance estimates and the coverage rate (CR) of the 95\% confidence intervals. The PS1 and PS3 estimators exhibited negligible bias across both study variables. In contrast, the PS2 estimators—based on the one-step GEC approach—showed noticeably larger biases. This bias stems from the use of equal propensity scores in the one-step GEC calibration, which can lead to biased estimates when the outcome model is misspecified.
However, the one-step GEC estimators achieved lower SE compared to other estimators, even the oracle estimators. This is because using estimated propensity score is more efficient than using the true propensity scores, as discussed in Section 7.3 of \citet{kimshao21}. 
Additionally, the choice of entropy measure did not significantly affect RMSE, with the SKL and EL estimators demonstrating nearly identical performance.

\begin{remark}
    Based on our findings, we recommend avoiding the use of the SKL entropy when some sampling rates are large ($d_i > 1 - 1/e \approx 0.63$), as it may lead to convergence issues during weight calibration. Additionally, when negative weights are not expected—as is typically the case in general survey settings—the SL entropy is also not recommended.
\end{remark}

\section{Real data analysis}
We compared the performance of the proposed method using two survey datasets: the 2022 Korea National Health and Nutrition Examination Survey (KNHANES) and the 2022 Korea Health Panel Survey (KHPS). The KNHANES is a national health and nutrition survey that has been conducted annually by the Korea Centers for Disease Control and Prevention since 2007. It aims to produce nationally representative statistics on a wide range of health indicators, including clinical measurements and dietary intake of the Korean population. 

Korean Health Panel Survey (KHPS), on the other hand, captures detailed usage patterns of healthcare services, in addition to comprehensive information on the health status and socio-demographic characteristics of the Korean population. Both KNHANES and KHPS use the two-stage stratified cluster sampling with survey districts and households as the primary and secondary sampling units. Furthermore, both surveys produced the sampling weights through nonresponse adjustment and postratification.

Using the KNHANES and KHPS datasets, we are interested in estimating the total number of individuals with hypertension across genders. While the KHPS has larger sample sizes($n = 13,799$) compared to the KNHANES($n = 6,265$), resulting in smaller standard errors, a high nonresponse rate of 33\% regarding chronic disease management has led to under-coverage in the samples. 
In this context, the KHPS can be considered a voluntary sample when analyzing the counts of individuals diagnosed with hypertension.

We considered three estimators: (1) the naive estimator using the complete cases from KHPS weighted by the design weights, (2) the HT estimator using KNHANES, and the one-step GEC estimator discussed in Section 3 using three entropies: (3) Kullback-Leibler(KL), (4) Hellinger distance(HD), and (5) Shifted Kullback-Leibler(SKL). For each version of the proposed GEC estimator, we included KHPS covariates relevant to the onset of hypertension, namely, sex, region, their interaction terms, age, and weight. The population totals for these covariates were estimated using the probability sample from KNHANES. The standard errors were computed using the \texttt{survey} package in \texttt{R}, by applying the calibrated GEC weights within the \texttt{svydesign} function.

Table \ref{realdata} summarizes the estimated domain totals and their standard errors. Despite the small standard error of the naive estimator, its point estimate is lower than that of the HT estimator, suggesting significant nonresponse bias.  The HT estimator, on the other hand, shows large variances due to the small sample size of KNHANES data. 
The proposed GEC estimator serves as a compromise between the naive and HT estimators, adjusting for nonresponse bias with a reasonably small standard error. The 95\% confidence intervals for the point estimates are presented in Figure \ref{95ci}. For the male group, the choice of entropy measures had little impact on the construction of the confidence intervals.

\begin{table}[ht]
\centering
\begin{tabular}{@{}ccrcrcrrrrr@{}}
\toprule
\multirow{2}{*}{}          & \multicolumn{2}{c}{\multirow{2}{*}{Naive}}        & \multicolumn{2}{c}{\multirow{2}{*}{HT}}           & \multicolumn{6}{c}{GEC}                                                                                                                                     \\ \cmidrule(l){6-11} 
                           & \multicolumn{2}{c}{}                              & \multicolumn{2}{c}{}                              & \multicolumn{2}{c}{KL}                            & \multicolumn{2}{c}{HD}                             & \multicolumn{2}{c}{SKL}                            \\ \midrule
                           & total                    & \multicolumn{1}{c}{SE} & total                    & \multicolumn{1}{c}{SE} & total                    & \multicolumn{1}{c}{SE} & \multicolumn{1}{c}{total} & \multicolumn{1}{c}{SE} & \multicolumn{1}{c}{total} & \multicolumn{1}{c}{SE} \\ \midrule
\multicolumn{1}{r}{Male}   & \multicolumn{1}{r}{3.96} & 0.14                   & \multicolumn{1}{r}{5.10} & 0.29                   & \multicolumn{1}{r}{4.92} & 0.15                   & 4.91                      & 0.14                   & 4.99                      & 0.13                   \\
\multicolumn{1}{r}{Female} & \multicolumn{1}{r}{4.27} & 0.13                   & \multicolumn{1}{r}{4.77} & 0.24                   & \multicolumn{1}{r}{4.74} & 0.12                   & 4.92                      & 0.12                   & 5.15                      & 0.12                   \\ \hdashline
\multicolumn{1}{r}{Total}  & \multicolumn{1}{r}{8.23} & 0.18                   & \multicolumn{1}{r}{9.88} & 0.45                   & \multicolumn{1}{r}{9.65} & 0.18                   & 9.83                      & 0.17                   & 10.14                     & 0.16                   \\ \bottomrule
\end{tabular}
\caption{Point estimates($\times 10^6$) and standard  errors($\times 10^6$) of domain totals for the prevalence of hypertension by gender.}
\label{realdata}
\end{table}

\begin{figure}[!ht]
    \centering
    \includegraphics[width=0.8\linewidth]{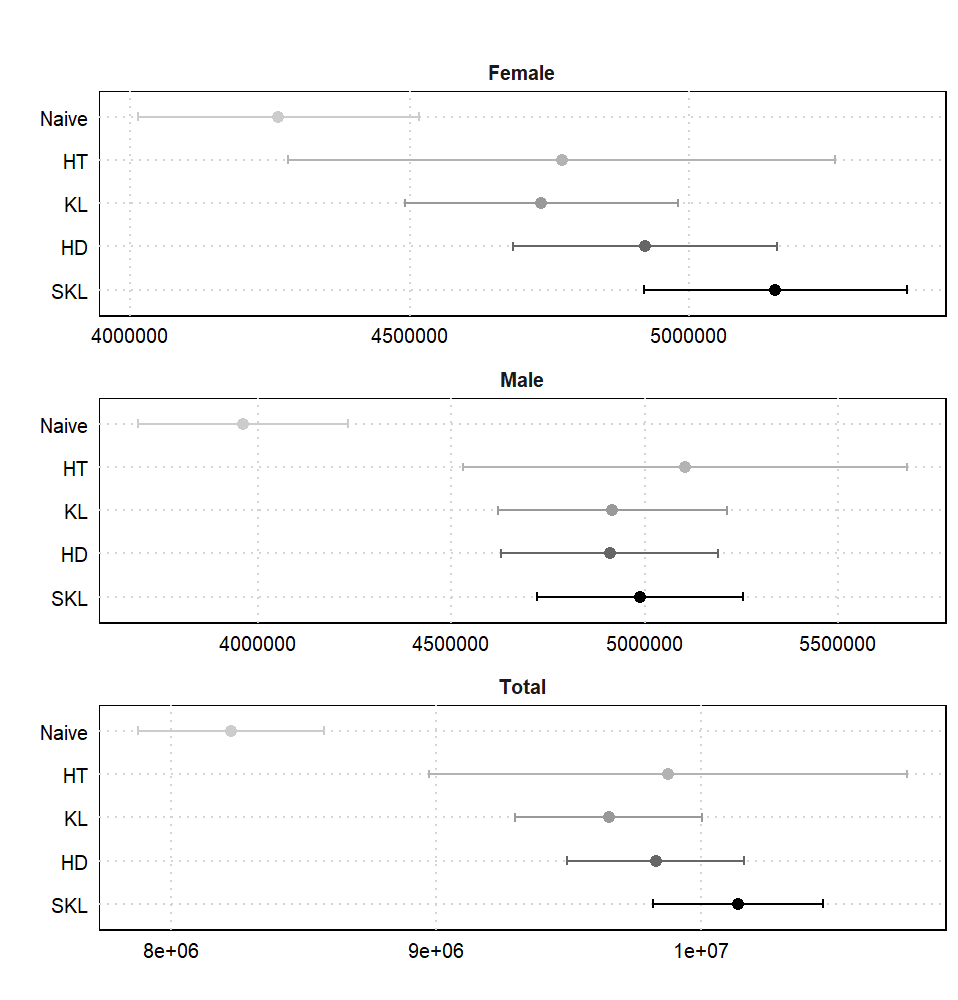}
    \caption{95\% Confidence intervals for the estimated domain totals for the prevalence of hypertension by gender.}
    \label{95ci}
\end{figure}

\section{Concluding Remark}

We have presented a unified framework for analyzing voluntary survey data using generalized entropy calibration (GEC). The dual relationship between the GEC estimator and the generalized regression estimator has been clarified to provide insight into the structure of calibration weighting. Fundamentally, the GEC approach implicitly incorporates the outcome regression model. Moreover, the choice of the generalized entropy function used for calibration determines the weighting scheme applied in the computation of the regression coefficients in the generalized regression estimator.

We have also developed a two-step calibration approach. The first step incorporates the propensity score (PS) model, while the second step uses covariates from the outcome regression model for calibration. Additionally, by including a debiasing calibration constraint, the resulting estimator is justified under the PS model, thereby achieving double robustness. The proposed methodology can be readily extended to construct multiply robust estimators, as in \citet{Han2014}.  
The method is implemented in the open-source \texttt{R} package \texttt{GECal}, available at \url{https://CRAN.R-project.org/package=GECal}. Moreover, the proposed approach is directly applicable to data integration problems involving non-probability and probability samples, as demonstrated in Section 6. 

In this paper, we have focused on settings where the number of covariates used for calibration is small or moderate. When the number of covariates becomes large, the asymptotic properties established in this work may no longer hold. However, leveraging the dual relationship between calibration and regression, regularization techniques commonly used in high-dimensional regression analysis can be incorporated into the calibration framework \citep{gao2023}. Further investigation in this direction will be pursued in future work.


\section*{Acknowledgements}

The research of the second  author was done during his visit to 
Seoul National University which was supported by Brain Pool program funded by the
Ministry of Science and ICT through the National Research Foundation of
Korea (RS-2023-00218474).  
His research was also partially supported by a grant from the U.S. National Science Foundation (2242820)  and a grant from the U.S. Department of Agriculture’s National Resources Inventory, Cooperative Agreement NR203A750023C006, Great Rivers CESU 68-3A75-18-504.
\section*{Appendix}

\setcounter{equation}{0}
\def\theequation{A.\arabic{equation}}

 \subsection*{Proof of Theorem 1}

 First, we write $\widehat{Y}_{\rm GEC} = \widehat{Y}_{\rm GEC} ( \hat{\blambda} ) $
 to  emphasize its dependency on $\hat{\blambda}$.  Now, define 
 \begin{equation}
 \widehat{Y}_{\ell} \left( \blambda , \bgamma \right) = \widehat{Y}_{\rm GEC} ( \blambda) + \left\{ \sum_{i=1}^N \bx_i  - \sum_{i=1}^N \delta_i \rho^{(1)} \left( \bx_i^{\top} \blambda \right) \bx_i \right\}^\top \bgamma . 
 \label{lin} 
 \end{equation}
Note that $\widehat{Y}_{\ell} \left( \blambda, \bgamma \right)$ satisfies the following two properties. 
\begin{enumerate}
\item $\widehat{Y}_{\ell} ( \hat{\blambda}, \bgamma ) = \widehat{Y}_{\rm GEC}$ for all $\bgamma$. 
\item Let $\blambda^*$ be the probability limit of $\hat{\blambda}$.  If $\bgamma^*$ is  the solution to 
\begin{equation}
 E \left\{ \nabla_{\lambda} \widehat{Y}_{\ell} \left( \blambda^*, \bgamma \right) \right\}= \mathbf{0}, 
 \label{randles}
 \end{equation}
then $\widehat{Y}_{\ell} ( \hat{\blambda}, \bgamma^*)$ is asymptotically equivalent to $\widehat{Y}_{\ell} ( {\blambda}^*, \bgamma^*)$ in the sense that 
$$ \widehat{Y}_{\ell} ( \hat{\blambda}, \bgamma^*) = \widehat{Y}_{\ell} ( {\blambda}^*, \bgamma^*) + o_p \left(n^{-1/2} N \right) 
$$
\end{enumerate}
Condition (\ref{randles}) is often called Randles' condition \citep{randles82}. For $\widehat{Y}_{\ell}$ in (\ref{lin}), the  equation for $\bgamma^*$ satisfying (\ref{randles}) 
can be written as 
\begin{equation}
 E \left\{ \sum_{i=1}^N \delta_i \rho^{(2)} \left( \bx_i^{\top} \blambda^* \right) \left( y_i - \bx_i^{\top} \bgamma^* \right) \bx_i \right\} = \mathbf{0} , 
 \label{randles2}
 \end{equation}
where 
$ \rho^{(2)} (\nu) = d^2  \rho ( \nu) / d \nu^2. $ Thus, $\bgamma^*$ is consistently estimated by 
$$ \hat{\bgamma}^* = \left( \sum_{i \in S} \hat{q}_i \bx_i \bx_i^{\top} \right)^{-1} \sum_{i \in S} \hat{q}_i 
\bx_i y_i,  $$
where 
 $\hat{q}_i=\rho^{(2)} \left( \bx_i^{\top} \hat{\blambda}  \right)$.

\subsection*{Proof of Theorem 2}

The generalized entropy calibration PS estimator is constructed in two-steps. In the first step, $\hat{\bm \phi}$ is computed by solving 
\begin{equation}
\hat{U}_1 \left( \bphi \right) \equiv  \sum_{i \in S} \rho_1^{(1)} ( \bx_{i}^{\top} \bphi ) \bx_{1i} - \sum_{i=1}^N \bx_{1i} = \mathbf{0} . 
\label{eq:24}
\end{equation}
In the second step, for a given $\hat{\bm \phi}$, $\hat{\bm \lambda}$ is computed by 
\begin{equation}
\hat{U}_2 \left( \hat{\bphi} , \blambda\right) \equiv  \sum_{i \in S} \rho_2^{(1)} \left( \hat{\bz}_{i}^{\top} \blambda \right)  \hat{\bz}_i   - \sum_{i=1}^N \hat{\bz}_i   = \mathbf{0} .
\label{eq:25}
\end{equation}
where  $\hat{\bz}_i = \left( \bx_{2i}^\top, \hat{g}_{2i} \right)^{\top}$ and  $\hat{g}_{2i} =  g_2 \left\{\rho_1^{(1)} (\bx_i^\top \hat{\bphi})\right\}$.  
  To properly reflect the uncertainty of the two estimated parameters, we apply Taylor expansion separately in the two-step calibration estimator.

First, for a given $\hat{\bm \phi}$, the generalized entropy calibration estimator can be expressed as  
$ \widehat{Y}_{\rm GEC2} = \widehat{Y}_{\rm GEC2} (  \hat{\bm \lambda} ), $
where $\hat{\bm \lambda}$ solves  (\ref{eq:25}). In this case, by \eqref{eq:12-res1}, 
we can express 
\begin{equation}
\widehat{Y}_{\rm GEC2}
= \widehat{Y}_{\rm GEC2, \ell 2}  + o_p \left(n^{-1/2} N \right) , 
\label{eq:12-res2}
\end{equation}
where 
 \begin{eqnarray*}
    \widehat Y_{\rm  GEC2,   \ell 2}  &=& \sum_{i=1}^N \left\{   {\bz}_i^\top  {\bm \gamma}_2^*    +   \delta_i \rho_2^{(1)}  \left( \bz_i^{\top} {\blambda}^* \right)   \left(   y_i - {\bz}_i^\top   {\bm \gamma}_2^*  \right)  
    \right\}   ,   \end{eqnarray*}
 ${\bz}_i^\top  =\left(\bx_{2i}^\top , \hat{g}_{2i} \right) $,   $\hat{g}_{2i} =g_2(  \hat{\omega}_{1i}) $, 
 $\blambda^* = p \lim \hat{\blambda}$, and ${\bm \gamma}_2^*$ is the probability limit of $\hat{\bm \gamma}_2$ in (\ref{gamma2}). 
 
Under the  PS model in (\ref{ps2}), using the same argument for proving Theorem 1 in \cite{lesage2019}, we can establish 
$$  \blambda^* = \left( \mathbf{0}^\top, 1 \right)^\top $$
so that 
$  \hat{\bz}_i^\top \blambda^*= g_2( \hat{\omega}_{1i})$  where $\hat{\omega}_{1i} = \rho_1^{(1)} ( \bx_{1i}^\top \hat{\bphi})$. Thus,  we have
 $$ \rho_2^{(1)}  \left( \hat{\bz}_i^{\top} {\blambda}^* \right)    =\rho_2^{(1)}  \left\{ g_2( \hat{\omega}_{1i}) 
 \right\} = \hat{\omega}_{1i} = \rho_1^{(1)} ( \bx_{1i}^\top \hat{\bphi})$$
and 
 \begin{eqnarray*}
    \widehat Y_{\rm  GEC2,   \ell 2} 
    &=&  \sum_{i=1}^N \left\{   \hat{\bz}_i^\top  {\bm \gamma}_2^*    +   \delta_i\rho_1^{(1)} ( \bx_{1i}^\top \hat{\bphi})\left(   y_i - \hat{\bz}_i^\top   {\bm \gamma}_2^*  \right)  
    \right\}  .  \end{eqnarray*}

Now, it remains to apply Taylor expansion on $\widehat{Y}_{\rm GEC2, \ell 2}$ with respect to $\bm \phi$. 
Note that we can express $\widehat{Y}_{\rm GEC2, \ell 2} = \widehat{Y}_{\rm GEC2, \ell 2} ( \hat{\bphi})$ to emphasize its dependency on $\hat{\bphi}$.  
Define 
$$ \widehat{Y}_{\rm GEC2, \ell}  (\bm \phi, \bgamma_1  ) = 
\widehat{Y}_{\rm GEC2, \ell 2} 
( {\bm \phi}  )   - \hat{U}_1 ( \bm \phi )^\top \bgamma_1 , $$
where $\hat{U}_1 (\bm \phi)$ is defined in (\ref{eq:24}), and note that 
\begin{equation}
 \widehat{Y}_{\rm GEC2, \ell} (\hat{\bm \phi},  \bgamma_1 ) = \widehat{Y}_{\rm GEC2, \ell 2} (\hat{\bm \phi})
 \label{eq:12-res2}
 \end{equation}
for all $\bgamma_1$. 
Our goal is to find $\bgamma_1^*$  such that 
 \begin{equation}
\widehat{Y}_{\rm GEC2, \ell} ( \hat{\bm \phi}, \bgamma_1^*  ) 
= \widehat{Y}_{\rm GEC2, \ell} ( {\bm \phi}^*,  \bgamma_1^* ) + o_p \left(n^{-1/2} N \right) 
\label{eq:12-res3}\end{equation}
 where $\bm \phi^* =  p \lim \hat{\bm \phi}$.  Under the correct PS model in (\ref{ps2}), we have $\bphi^* = \bphi_0$. 
 
 A sufficient condition for (\ref{eq:12-res3}) is 
\begin{equation}
 E\left\{ \frac{\partial}{ \partial \bphi} \widehat{Y}_{\rm GEC2, \ell} ( \bm \phi^*, \bgamma_1^* ) \right\}= \mathbf{0}  
 \label{eq:randles3}
 \end{equation}
which is the Randles' condition applied to 
$\widehat{Y}_{\rm GEC2, \ell} ( \bm \phi,   \bgamma_1 )$. Now, note that  
\begin{eqnarray} 
E\left\{ \frac{\partial}{\partial \bm \phi} 
\widehat{Y}_{\rm GEC2, \ell} ( \bm \phi^*, \bgamma_1  )\right\}
&=& E \left\{  \sum_{i=1}^N \delta_i \rho_1^{(2)} ( \bx_{1i}^\top \bphi^*)  \left( y_i- \bz_i^{\top} {\bgamma}_2^* - \bx_{1i}^{\top} \bgamma_1  \right) \bx_{1i}  \right\} \notag  \\
&+& E \left[ \sum_{i=1}^N  \dot{\bz}_i^\top \bm \gamma_2^* \left\{ 1  - \delta_i \rho_1^{(1)} ( \bx_{1i}^{\top} \bm \bphi^* ) \right\}   
\right] \label{eq:a.8}
\end{eqnarray} 
where $\dot{\bz}_i = \partial \bz_i / \partial \bm \phi$. By (\ref{adjust}), $\rho_1^{(1)}( \bx_i^{\top} \bm \bphi^* ) = 1/ \pi( \bx_i^\top \bm \phi^*)$. Also, under the PS model (\ref{ps2}),  $\bm \phi^* = \bm \phi_0$ and the second term of (\ref{eq:a.8}) is equal to zero under the PS model (\ref{ps2}). 
Thus, 
we obtain  
$$
\bgamma_1^* = p \lim \left( 
\sum_{ i \in S}  \rho_1^{(2)} ( \bx_{1i}^\top \bphi^* ) \bx_{1i} \bx_{1i}^{\top} 
\right)^{-1}  \sum_{i \in S} \rho_1^{(2)} ( \bx_{1i}^\top \bphi^* )  \left( y_i- \bz_i^{\top} \bgamma_2^*  \right) \bx_{1i}  $$
as the solution to (\ref{eq:randles3}). 
Therefore, combining (\ref{eq:12-res1}), (\ref{eq:12-res2}), and (\ref{eq:12-res3}), we can establish (\ref{result2}).

\bibliographystyle{apalike}  
\bibliography{reference}

\end{document}